\documentstyle[times,pramana,epsf,floats]{ias}
\newcommand{\beq}{\begin{equation}}
\newcommand{\eeq}{\end{equation}}
\newcommand{\beqarray}{\begin{eqnarray}}
\newcommand{\eeqarray}{\end{eqnarray}}

\def\lsim{\raise0.3ex\hbox{$\;<$\kern-0.75em\raise-1.1ex\hbox{$\sim\;$}}}
\def\gsim{\raise0.3ex\hbox{$\;>$\kern-0.75em\raise-1.1ex\hbox{$\sim\;$}}}

\def\erg{\,{\rm erg}}
\def\ergs{\,{\rm ergs}}
\def\ev{\,{\rm eV}}

\def\sec{\,{\rm sec}}

\begin{document}
\mark{{UHE Cosmic Rays and TeV Photons from Gamma Ray Bursts}{P 
Bhattacharjee}}
\title{Ultrahigh Energy Cosmic Rays and Prompt TeV Gamma Rays from Gamma 
Ray Bursts}
\author{Pijushpani Bhattacharjee$^1$ and Nayantara Gupta$^2$}
\address{$^1$Indian Institute of Astrophysics, Koramangala, Bangalore 
560034\\
$^2$Indian Association for the Cultivation of Science, Jadavpur, Kolkata 
700032} 
\keywords{Gamma ray bursts, Ultrahigh energy cosmic rays, TeV gamma rays}
\pacs{98.70.Rz, 98.70.Sa}
\abstract{Gamma Ray Bursts (GRBs) have been proposed as one 
{\it possible} class of sources of the Ultrahigh Energy Cosmic 
Ray (UHECR) events observed up to energies $\gsim10^{20}\ev$. 
The synchrotron radiation of the highest energy protons accelerated within 
the GRB source should produce gamma rays up to TeV energies. 
Here we briefly discuss the implications on the energetics of 
the GRB from the point of view of the detectability of the prompt TeV 
$\gamma-$rays of proton-synchrotron origin in GRBs in the up-coming 
ICECUBE muon detector in the south pole.}
\maketitle
\section{Introduction}
The origin of the observed Ultrahigh Energy Cosmic Ray  
(UHECR) events with estimated energy in excess of $10^{20}\ev$ 
\cite{uhecr_obs} is unknown and is currently a subject of much 
discussions~\cite{uhecr_revs}. It is 
in general extremely difficult to accelerate particles to such high 
energies in most of the known astrophysical objects through conventional
acceleration mechanisms. Gamma Ray Bursts (GRBs) have been 
proposed~\cite{grb_uhecr} as one {\it possible} class of sources which may 
be capable of accelerating particles to the requisite energies. 
If so, then synchrotron radiation of these protons in the magnetic 
field within the GRB source should produce prompt GeV -- TeV gamma 
rays~\cite{tevgrb} whose 
detection may provide important clues as to the nature of GRBs. The 
predicted photon number flux at TeV energies is generally 
too low to be detected by the satellite-borne detectors which have limited 
sizes. However, ground-based detectors can in
principle detect TeV photons from GRBs by detecting the secondary
particles comprising the ``air showers'' generated by
these photons in  the Earth's atmosphere.

Indeed, several ground-based gamma ray detectors employing different 
detection techniques~\cite{tev_grb_ground_detect} have independently 
claimed evidence, albeit not with strong statistical significance, for
possible TeV $\gamma$-ray emission from sources in directional and
temporal coincidence with some GRBs detected by BATSE. The estimated
energy in TeV photons have been generally found to be significantly 
larger (by up to 4 orders of magnitude in some cases) than the 
corresponding sub-MeV energies measured by BATSE. Note that, since TeV 
photons are efficiently absorbed in the intergalactic infrared (IR) 
background due to pair production~\cite{stecker-dejager}, only few 
relatively close by (i.e., low redshift) GRBs can be expected to be 
observed at TeV energies. 

The high (GeV--TeV) energy component of proton-synchrotron 
origin we are considering here is distinct from and not a continuation 
of the low energy (keV--MeV) component, the latter being due to 
synchrotron radiation of the electrons accelerated 
along with the protons in the same magnetic field within the GRB source.  
If the fundamental source of energy of the GRBs is indeed the
kinetic energy of the ultrarelativistic bulk flow of matter as in the
currently popular fireball model~\cite{piran_revs}, then, at least 
initially, one would expect the total energy content in protons to be 
higher than that in electrons by a factor of $\sim m_p/m_e\sim2000$, where 
$m_p$ and $m_e$ are proton and electron rest mass, respectively. The 
energy transfer from protons to electrons through coulomb interaction is 
an inefficient process~\cite{totani_coulomb}. Immediately after 
dissipation of the kinetic energy of the bulk 
flow through formation of internal shocks, the total energy 
content of protons would, therefore, be higher than that of electrons.
If the proton spectrum is sufficiently hard so that most of the 
total energy in the proton component lies at the highest energy end of the 
spectrum where the synchrotron emission process is efficient, and if the 
pair-production process for the resulting TeV synchrotron photons on the 
ambient low-energy photons within the GRB source is inefficient, then the
bulk of the energy in the proton component (which is initially higher
than that in the electron component) will escape from the source in
the form of TeV photons, giving significantly
higher total energy in the TeV photon component compared to that in the
sub-MeV component. 
\section{Results and Discussions} 
We have recently studied~\cite{bg} the detectability of the possible TeV 
component of proton-synchrotron origin from GRBs in the up-coming ICECUBE 
muon  
detector~\cite{icecube} in the south pole which can 
detect TeV energy photons by detecting the muons produced by TeV photons 
in the Earth's atmosphere~\cite{icecube_advantage}. In Figure 1 we 
show (see for details Ref.~\cite{bg})  
the behavior of the minimum luminosity (assuming isotropic emission) in 
the high energy component required for detection with a signal to noise 
ratio of 5 or 
larger in the ICECUBE detector, as a function of $\alpha_p$, the 
differential spectral index of the accelerated protons within the GRB, 
for various values of the redshift $z$ of the GRB. 
The calculations include the 
effects of both the internal (i.e., within the GRB environment) as well as 
the external (i.e., in the intergalactic medium) optical depth of TeV 
photons. The required isotropic
luminosity of the high energy photon component is upward of
$10^{56}\ergs/\sec$ (for reasonably hard proton spectrum) and are 
generally more than 3--4 orders
of magnitude higher than typical estimated isotropic luminosities 
($\sim10^{53}\ergs/\sec$) in the keV--MeV BATSE energy band. 

In order to explain the observed
flux of UHECR, one needs a typical GRB to emit a total energy of $\sim
{\rm few\, } \times 10^{53}\erg$ (assuming isotropic emission) in UHE 
protons with energy $>10^{19}\ev$\cite{bahcall-waxman_uhecr}. Note that
the above number refers to the
total energy {\it escaping} from the GRB source in the form of UHE
protons. 
\begin{figure}[htbp]
\epsfxsize=8cm
\centerline{\epsfbox{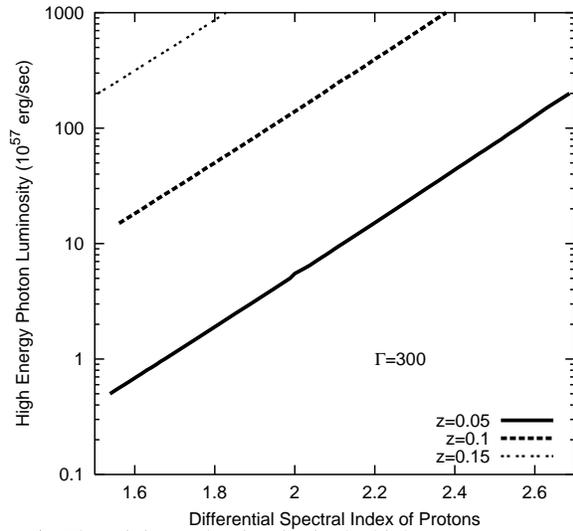}}
\caption{The minimum luminosity in the high energy component 
required for detection with a signal to noise ratio of 5 or larger in a
ICECUBE class detector, as a function
of the power-law index ($\alpha_p$) of the differential spectrum of
protons accelerated within a typical GRB source, for various values of the
redshift $z$ of the GRB. The values taken for other relevant parameters are 
as given in Ref.~\protect\cite{bg}.}
\label{fig:req_lum}
\end{figure}
The total energy in the UHE proton component produced {\it
within} the GRB may be significantly higher, depending on the various
energy loss processes of the UHE protons within the GRB source. If
synchrotron radiation is the dominant energy loss process for the UHE
protons within the GRB source, and if the pair-production optical depth of 
the resulting GeV--TeV
photons within the source is sufficiently small, then the escaping
GeV--TeV photon luminosity of the source can be larger than the
escaping UHE proton luminosity; see Ref.~\cite{bg} for conditions on 
various parameters under which this can happen.

\end{document}